\documentclass[10pt]{article}
\usepackage[centertags]{amsmath}
\usepackage{amsfonts}
\usepackage{graphicx}
\usepackage{amssymb}
\newcommand{\id}{i\kern.06em\hbox{\raise.25ex\hbox{$/$}\kern-.60em$\partial$}}
\newcommand{\dd}{\kern.06em\hbox{\raise.25ex\hbox{$/$}\kern-.60em$\partial$}}
\newcommand{\beq}{\begin{equation}}
\newcommand{\eeq}{\end{equation}}
\newcommand{\ba}{\begin{eqnarray}}
\newcommand{\ea}{\end{eqnarray}}
\newcommand{\be}{\begin{equation}}
\newcommand{\ee}{\end{equation}}
\newcommand{\bea}{\begin{eqnarray}}
\newcommand{\eea}{\end{eqnarray}}

\usepackage[centertags]{amsmath}
\usepackage{amsfonts}

\begin{document}
\title{BPS Equations and the Stress Tensor}
\author{E.~F.~Moreno$^a$,
F.~A.~Schaposnik$^b$\thanks{Associated with CICBA}
\\
\\
{\normalsize $^a\!$\it Department of Physics,West Virginia University}\\
{\normalsize\it Morgantown, West Virginia 26506-6315, U.S.A.}
\\
{\normalsize $^b\!$\it Departamento de F\'\i sica, Universidad
Nacional de La Plata}\\ {\normalsize\it C.C. 67, 1900 La Plata,
Argentina}}
\maketitle
\begin{abstract}
We exploit the relationship between the space components
$T_{ij}$ of the energy-momentum tensor and the supercurrent to
discuss the connection between the BPS equations and the
vanishing of the components of the stress tensor in various
supersymmetric theories with solitons.
Using the fact that certain combination of supercharges
annihilate BPS states, we show that $T_{ij}=0$ for kinks,
vortices and dyons, displaying the connection between
supersymmetry and non-interacting BPS solitons.
\end{abstract}

\section*{Introduction}
First order BPS equations were originally obtained either by looking
for a bound of the soliton mass \cite{Bogo} or by imposing the
stress tensor  to vanish \cite{dVS}. Already in this last work the
relation between supersymmetry and the possibility of reducing the
second order equations of motion to BPS equations at certain
critical values of the coupling constants was stressed and
afterwards exploited in the search of classical solutions to two
dimensional supersymmetric models \cite{DiVF}.

The origin of such a connection was finally clarified by Witten
and Olive  \cite{WO} by considering the supersymmetric
extension of bosonic models {  exhibiting  topological} soliton
solutions. Studying the supersymmetry algebra,  it was shown in
this work that the soliton topological charge can be identified
with the central charge of the supercharge algebra  and gives a
lower bound for the soliton mass. This was done  for the
supersymmetric version of a scalar field theory in $1+1$
dimensions with kink solutions and a ${\cal N} = 2$ Yang-Mills
theory in $3+1$ dimensions with dyon solutions. Afterwards, the
case of vortices in ${\cal N} = 2$ supersymmetric gauge
theories in $2+1$ dimensions and instantons in 4 dimensional
Euclidean space was discussed along the same lines
\cite{LLW}-\cite{ENS1} and the extension to the case of
supergravity models was also studied\cite{ENS2}. The question
on how supersymmetry protects the Bogomol'nyi bound at the
quantum level also  deserved a lot of attention
\cite{RvNW}-\cite{SY}.

We show in the present note how the alternative derivation of
BPS equations from the vanishing of the soliton stress-tensor
$T_{ij}$ ($i,j=1,2,3$) can be also understood supersymmetry
point of view studying the supercurrent-supercharge algebra. As
it is well known, the algebra of supersymmetry itself already
imposes a very intimate relationship between the supercurrent
and the stress tensor. This relationship stems from the
connection between the energy-momentum vector and the
supercharge \cite{Sohnius}. In fact both the supercurrent and
$T_{\mu\nu}$  must belong to the same supermultiplet and then
it is not difficult   to understand how an identity between the
stress tensor and an appropriate trace containing the
supersymmetric transform of the supercurrent connects BPS
states and the condition
\be \langle BPS| T_{ij}|BPS\rangle = 0 \; . \label{base} \ee

In order to construct the supercurrent and show how its connection
with the energy-momentum tensor leads to eq.(\ref{base})  we will
work with specific models having BPS $1+1$ dimensional kinks and
$2+1$ vortices and also explain how the results can be easily
extended to the case to BPS dyons in $3+1$ dimensions. In fact our
derivation indicates that the same result should be also valid for
any other model with BPS soliton solutions.

It should be mentioned that our work was prompted by a recent work
of Manton \cite{Manton} where new scaling identities for solitons
are derived in terms of the stress tensor, showing the relevance of
$T_{ij}$  in connection with the study of soliton solutions. As
mentioned above, already in the case of vortices it was recognized
\cite{dVS} that the critical point at which the topological bound
for the energy of the Abelian Higgs vortices is saturated
corresponds to the limiting value between type-I and type-II
superconductivity, precisely where forces between vortices (and
hence the surface integral of $T_{ij}$) vanish \cite{Abri}. We shall
see below that supersymmetry provides a way to construct models
where general noninteracting solitons equations can be studied by
analyzing the Noether supercurrent.

\section*{Scalar field theory in two dimensions}
The action for the simplest two-dimensional supersymmetric
model admitting solitons in its bosonic sector reads, in
component fields \cite{dAdV},
\be
S= \int d^2x \left( \frac{1}{2} \partial_\mu \phi
\partial^{\mu}\phi + \frac{i}{2}\, \bar \psi \dd \psi +
\frac{1}{2} F^2 + F\,V[\phi] -\frac{1}{2}\, V'[\phi]\, {\bar \psi} \psi \right)
\label{uno}
\ee
with $\phi$ a real scalar field, $\psi$ a 2-component Majorana
spinor, F  an auxiliary field and $V[\phi]$   an arbitrary
function. We take the metric $g_{\mu\nu}$ with signature
$(+,-)$ and the Dirac matrices as
\[ \gamma^0 =\sigma_2 =
\begin{pmatrix}
0&-i\\
i&0
\end{pmatrix}
\;\;\;\;\; \gamma^1 =i \sigma_1 =
\begin{pmatrix}
0&i\\
i&0
\end{pmatrix}
\]
With this conventions the charge conjugation matrix satisfying
$C\cdot \gamma^{\mu}\cdot C^{-1} = -\gamma^{\mu}$ is given by
$C=-\gamma^0$. Given a spinor
\[
\psi = \begin{pmatrix}
\psi_+\\
\psi_-
\end{pmatrix}\;\;, \;\;\; {\bar \psi} = \psi^{\dagger}\, \gamma^0
\]
the charge conjugate $\psi^c$ is then
\[\psi^c = C\, {\bar
\psi}^T = \psi^*\]
 so that   $\psi_+$ y $\psi_-$ are real and
\[ \bar \psi = i\left( \psi_-, - \psi_+\right) \]
The energy momentum tensor associated with action (\ref{uno})
takes the form
\begin{align}
T_{\mu \nu} = \partial_\mu \phi \partial_\nu \phi +
\frac{i}{2}\bar \psi \gamma_{\mu}\partial_{\nu} \psi
-\frac{1}{2}\, g_{\mu \nu} \left(\partial_\alpha \phi
\partial^\alpha \phi -  V^2
+ i\,\bar \psi \gamma^\alpha \partial_\alpha \psi -
V'\, \bar \psi \psi \right)
\end{align}
and its symmetric on-shell components
\begin{align}
T_{00} &= \frac{1}{2}\left( (\partial_0 \phi)^2 + (\partial_1
\phi)^2\right) +  \frac{1}{2} V^2 + \frac{1}{2} V'\, {\bar \psi}\psi
- \frac{i}{2} {\bar \psi} \gamma^1\partial_1 \psi \nonumber\\
T_{11} &= \frac{1}{2}\left( (\partial_0 \phi)^2 + (\partial_1
\phi)^2\right) - \frac{1}{2} V^2 - \frac{1}{2} V'\, {\bar \psi}\psi
+ \frac{i}{2} {\bar \psi} \gamma^0\partial_0 \psi\\
T_{01} &= \partial_0\phi \partial_1\phi + \frac{1}{2}\,
\psi \gamma^0\partial_1 \psi\\
T_{10} &= \partial_0\phi \partial_1\phi - \frac{1}{2}\,
\psi \gamma^1\partial_0 \psi
\end{align}

The (off-shell) supersymmetric transformations leaving
action(\ref{uno}) invariant are
\begin{align}
\delta \phi &= {\bar \epsilon} \psi\nonumber\\
\delta \psi &= - i \, \dd \phi\, \epsilon + F\, \epsilon\nonumber\\
\delta F &= - i \, {\bar \epsilon}\, \dd \psi
\label{dos}
\end{align}
and the associated conserved supercurrent is
\be
 J_{\mu} = (\dd \phi + i\, V)\,
\gamma_{\mu} \psi
\ee
where the auxiliary field has been eliminated using its
equation of motion. More explicitly
\begin{align}
J_0 &= ~ (\dd \phi + i\, V)\,\gamma^0 \psi   =
\begin{pmatrix}
~ (\partial_- \phi) \psi_+ + V\, \psi_-\\
~ (\partial_+ \phi) \psi_- - V\, \psi_+
\end{pmatrix}\; , \nonumber\\
 J_1 &= \!\!-(\dd \phi + i\, V)\,\gamma^1 \psi =
\begin{pmatrix}
-(\partial_- \phi) \psi_+ + V\, \psi_-\\
(\partial_+ \phi) \psi_- + V\, \psi_+
\end{pmatrix}
\label{jotas}
\end{align}
The chiral components $Q^{\pm}$ of the supersymmetry charge
take then the form
\begin{align}
Q_+ &= \int dx\; \left\{(\partial_- \phi) \psi_+ + V\, \psi_-\right\}
\nonumber\\
 Q_- &= \int dx\; \left\{ (\partial_+ \phi) \psi_- - V\, \psi_+
\right\}
\label{ques}
\end{align}
with
$\partial_\pm = \partial_0 \pm \partial_1$.
Concerning $\bar Q$, one has
$\bar Q= Q^\dagger\, \gamma^0 = \left( i \, Q_-, -i\, Q_+\right)$.

The equal-time commutations/anti-commutation  relations
are
\begin{align}
  [\phi(x), \partial_0 \phi(x')] &= i\, \delta(x-x')\nonumber\\
 \{\psi_+(x), \psi_+(x')\} &= \delta(x-x')\nonumber\\
 \{\psi_-(x), \psi_-(x')\} &= \delta(x-x')
 \label{cro}
\end{align}
From this, one finds for   the supercharge algebra (in the
rest frame)
\be \{Q_\pm,Q_\pm\} = 2(M \pm Z) \label{cargas}
\ee
where
\be M = \int dx\, T_{00} \ee
Concerning  $Z$, it is given by
\be Z = \int dx\, V[\phi]\frac{\partial\phi}{\partial x} = \int
dx\, \frac{\partial W}{\partial x} \ee
and coincides with the topological charge which is non-trivial
for soliton states.

In order to find the Bogomol'nyi bound, Witten and Olive
considered \cite{WO}  the combinations
\begin{align}
Q_+ + Q_- &= \int\left\{ \left( \partial_+ \phi + V\right)\psi_-
+  \left( \partial_- \phi - V\right)\psi_+\right\} \\
Q_+ - Q_- &= \int\left\{ -\left( \partial_+ \phi - V\right)\psi_-
+  \left( \partial_- \phi + V\right)\psi_+\right\}
\end{align}
Then, writing
\begin{align}
2M &= ~~Z + (Q_+ + Q_-)^2 \nonumber\\
2M &= -Z + (Q_+ -Q_-)^2
\end{align}
one finds  that the soliton mass $M$ is bounded by the topological charge,
\be
M \geq \frac{|Z|}{2}
\ee
and that the bound is attained for those states $|BPS\rangle_\pm$ such that
\be (Q_+ + Q_-)|BPS\rangle_+ = 0 \label{14a} \ee
or
\be (Q_+ - Q_-)|BPS\rangle_- = 0 \label{15a} \ee
In view of the explicit form of charges these states correspond
to kink  solutions satisfying the first order BPS equations
\begin{align}
&\partial_0\phi = 0\; , \;\;\;\; \partial_1\phi =
V\;\;\;\;\;\;\;\;\;  +\text{kink} \\
&\partial_0\phi = 0\; , \;\;\;\; \partial_1\phi =
-V\;\;\;\;\;\; -\text{anti-kink}
\end{align}
which can be written in the form
\begin{align}
&\partial_+\phi -V= 0\; , \;\;\;\; \partial_-\phi + V=
0\;\;\;\;\;\; +\text{kink} \\
&\partial_+\phi +V =0\; , \;\;\;\; \partial_-\phi  -V=
0\;\;\;\;\;\; -\text{anti-kink}
\end{align}
Each of the BPS kink solutions break half of the supersymmetry
of the theory according to the choice among eqs.(\ref{14a}) or
(\ref{15a}).

Let us now study the supercurrent-supercharge anticommutators.
In particular, from the canonical commutation relations
(\ref{cro}) one has
\be \{J^\mu_\alpha,\bar Q_\beta\}
= 2i\gamma^\rho_{\alpha\beta} T^{\;\mu}_\rho + 2
i{\gamma_5}_{\alpha\beta} \xi^\mu
\label{shi} \ee
with $J^\mu$ the supercurrent and $\xi^\mu$ the topological
current,
\be \xi^\mu = V\, \epsilon^{\mu \nu}\partial_\nu \phi
\ee
related to the central charge through the identity
\be
\int dx \xi^0 = Z
\ee
Writing
\be \mathcal{M}_{\alpha \beta}  = \{J^1_\alpha,\bar Q_\beta\} \ee
one easily finds
\begin{align}
\{\gamma^1, \mathcal{M}\} &= 2 \{\gamma^1, \gamma^\nu\}\,
T^1_\nu + 2 i\{\gamma^1, {\gamma_5} \}\, \xi^\mu\\
&=- 4\, T_{11}
\end{align}
Explicitly, the l.h.s. takes the form
\be \{\gamma^1, \mathcal{M}\} =
\begin{pmatrix}
\{J_{1-}, Q_-\} - \{J_{1+},Q_+\} & \{J_{1+}, Q_-\} -
\{J_{1-},Q_+\} \\
\{J_{1+}, Q_-\} - \{J_{1-},Q_+\} & \{J_{1-}, Q_-\} -
\{J_{1+},Q_+\}
\end{pmatrix}
\ee
and then
\begin{align}
&\{J_{1-}, Q_-\} - \{J_{1+},Q_+\} = 4\, T_{11}\\
&\{J_{1+}, Q_-\} - \{J_{1-},Q_+\} = 0
\end{align}
From these two equations, one can write two identities for the
stress-tensor
\be
T_{11} = -\frac{1}{4}\{J_{1+}+J_{1-}, Q_+ - Q_- \}
\label{ec-1}
\ee
\be T_{11}=   -\frac{1}{4}  \{J_{1+}-J_{1-}, Q_+ + Q_- \}
\label{ec-2}\ee
Then,  in view of (\ref{14})-(\ref{15}) and being the  currents
$J_{1\pm}$ given by
\begin{align}
J_{1+} + J_{1-} &= \left( \partial_+ \phi + V\right)\psi_-
-  \left( \partial_- \phi - V\right)\psi_+\\
J_{1+} - J_{1-} &= -\left( \partial_+ \phi - V\right)\psi_-
-  \left( \partial_- \phi + V\right)\psi_+
\end{align}
we conclude that either
\be
_+\langle BPS |T_{11}|BPS\rangle_+ = 0
\ee
or
\be
 _-\langle BPS |T_{11}|BPS\rangle_- = 0
\ee
That is, BPS saturated states preserving half of the
supersymmetry correspond to states with vanishing stress
tensor.

\section*{Scalar QED in three dimensions}

Our conventions for $\gamma$-matrices,
(${\gamma^{\mu})_{\alpha}}^{\beta}$ are,
\begin{equation}
\gamma^0 = \left( \begin{array}{rr} 0 & -i \\ i & 0 \end{array}
\right) \; \; \; \; \; \gamma^1 = \left( \begin{array}{rr} 0 & i \\
i & 0 \end{array} \right) \; \; \; \; \; \gamma^2 = \left(
\begin{array}{rr} i & 0 \\ 0 & -i \end{array} \right) \label{5}
\end{equation}
\[ \gamma^{\mu}\gamma^{\nu} = g^{\mu\nu} +
i\epsilon^{\mu\nu\lambda}\gamma_{\lambda}  \]
with the metric with signature $(+ - -)$.

The  ${\cal N}=2$ supersymmetric action associated with the
Abelian Higgs model is
\begin{align}
{\cal S}_{N=2} & =  \int d^3x \{ -\frac{1}{4}F_{\mu\nu}F^{\mu\nu}
+ \frac{1}{2}(\partial_{\mu}N)(\partial^{\mu}N) + \frac{1}{2}
(D_{\mu}\phi)^*(D^{\mu}\phi) - \frac{e^2}{4}N^2\vert\phi\vert^2 \nonumber \\
& \;\;\;\;-  \frac{e^2}{8}(\vert\phi\vert^2 - {\phi_0}^2)^2 +
\frac{i}{2}\overline{\Sigma}\not\!\partial\Sigma +
\frac{i}{2}\overline{\psi}\not\!\! D\psi -
\frac{e}{2} N\overline{\psi}\psi \nonumber \\
& \;\;\;\;-  \frac{e}{2}(\overline{\psi}\Sigma\phi + h.c.) \label{10bis}
\end{align}
Concerning boson fields,  $A_\mu$ is an Abelian gauge field,
$F_{\mu\nu}$ its curvature, $\phi$ a complex scalar and  $N$ a
real scalar field. The covariant derivative is defined as \be
D_\mu\phi = \partial_\mu \phi + i e A_\mu \ee Note that the
coupling constant in the gauge symmetry breaking scalar potential
is  taken as $\lambda = e^2/8$, the condition required in order to
have $N=2$ supersymmetry. Fermion fields $\psi$ and $\Sigma$  are
Dirac fermions and
\be \not\!\! D\psi  = (i \not\!\partial - e
\not\!\! A ) \psi
\ee
The energy-momentum tensor components of the bosonic sector are
\begin{align}
T_{ij}&= \left(\frac{1}{2}\,B^2  - \frac12\, |D_i\phi|^2 -
\frac{e}{8}^2 \left(|\phi|^2
- \phi_0^2\right)^2 \right) \delta_{ij}
+ \frac12\left(D_i\phi\right)^*\! D_j \phi +\frac12
\left(D_j\phi\right)^*\!D_i \phi \nonumber\\
T_{00} &= \frac{1}{2}\, B^2 + \frac{1}{2}\, \vert D_i\phi\vert^2
+ \frac{e^2}{8}(\vert\phi\vert^2 - {\phi_0}^2)^2
\label{cuatres}
\end{align}
where $B=F_{12}$.

Action (\ref{10bis}) is invariant under the following ${\cal
N}=2$ supersymmetry transformations
\begin{align}
 &{\delta} A_{\mu} = -i\overline{\eta}_c\gamma_{\mu}
\lambda &
 {\delta} \phi &= \overline{\eta}_c\psi
\nonumber\\
 &{\delta} \psi = \ -i\gamma^{\mu} D_{\mu}\phi \eta_c-
(8\lambda)^{1/2}N\phi \eta_c & {\delta} N &=
\overline{\eta}_c\chi \nonumber
\\
&{\delta} \Sigma =
- \left(\frac{1}{2}\epsilon^{\mu\nu\lambda}F_{\mu\nu}
\gamma_{\lambda} + (2\lambda)^{1/2}(\vert\phi\vert^2 - {\phi_0}^2)
+ i\not\!\partial N\right) \eta_c \label{9}
\end{align}
with $\eta_c$ a complex (Dirac spinor) parameter.

The Noether supercurrent associated with invariance of action
(\ref{10bis}) under transformations (\ref{9}) is
\begin{align}
{ \cal J}^\mu =& \bar\eta_c \left(
-\frac{1}{2}\epsilon^{\mu\nu\lambda} F_{\mu\nu}\gamma_{\lambda} +
i\not\!\partial N - \frac{e}{2}
(\vert\phi\vert^2 - {\phi_0}^2) \right) \gamma^\mu\Sigma \nonumber\\
&+   \bar\eta_c  \left( i(\not\!\! D\phi)^* - \frac{e}{2}N\phi^* \right)
\gamma^\mu \psi  +  \overline{\psi}\gamma^\mu \left( -i\not\!\! D\phi -
\frac{e}{2}N\phi \right) \eta_c \nonumber\\
&+\overline{\Sigma}\gamma^\mu\left(-
\frac{1}{2}\epsilon^{\mu\nu\lambda}F_{\mu\nu}\gamma_{\lambda} -
i\not\!\partial N - \frac{e}{2}(\vert\phi\vert^2 - {\phi_0}^2)
\right)  \eta_c
\end{align}
so that the conserved charge ${\cal Q}$ can be defined as
\begin{equation}
{\cal Q} = \frac{1}{\sqrt{2}\,e\phi_0}\int d^2x {\cal  J} ^0
\label{12}
\end{equation}
  Writing
\begin{equation}
{\cal Q} = \overline{\eta}_cQ + \overline{Q}\eta_c
\label{13}
\end{equation}
one finds
\begin{align}
Q & =  \frac{1}{\sqrt{2}\,e\phi_0}\int d^2x \left[\left(
-\frac{1}{2}\epsilon^{\mu\nu\lambda} F_{\mu\nu}\gamma_{\lambda} +
i\not\!\partial N - \frac{e}{2}
(\vert\phi\vert^2 - {\phi_0}^2) \right) \gamma^0\Sigma \right.
\nonumber \\
& \left. \phantom{\frac12}+  \left( i(\not\!\! D\phi)^* -
\frac{e}{2}N\phi^* \right)
\gamma^0 \psi \right] \label{14}
\end{align}
and
\begin{align}
\overline{Q}  = & \frac{1}{\sqrt{2}\,e\phi_0}\int d^2x \left[
\overline{\Sigma}\gamma^0\left(-
\frac{1}{2}\epsilon^{\mu\nu\lambda}F_{\mu\nu}\gamma_{\lambda} -
i\not\!\partial N - \frac{e}{2}(\vert\phi\vert^2 - {\phi_0}^2)
\right) \right. \nonumber \\
& \left. \phantom{\frac12} + \overline{\psi}\gamma^0
\left( -i\not\!\! D\phi -
\frac{e}{2}N\phi \right) \right] \label{15}
\end{align}
One can now compute the supersymmetry algebra among supercharges $Q$
and $\bar Q$. Since this will be connected with the Bogomol'nyi
bound for the Abelian Higgs model, we shall put $N=0$ and, after
using fermion anticommutator relations we shall also put all
fermions to zero. As one is interested in static configurations with
finite energy one should also impose $A_0$. The answer is
\begin{equation}
\{Q_{\alpha},\overline{Q}^{\beta}\} = {(\gamma_{0})_{\alpha}}^{\beta}P^{0}
+ {\delta_{\alpha}}^{\beta} Z
\label{16}
\end{equation}
where
\begin{equation}
P^0 =   \frac{1}{e^2{\phi_0}^2}
\int d^2x T_{00} \equiv M
\label{17}
\end{equation}
and the central charge $Z$ is given by
\begin{equation}
Z = \frac{1}{2\,e^2{\phi_0}^2}\int d^2x
\left[ e\, B(\vert\phi\vert^2 - {\phi_0}^2)
+ i\epsilon^{ij}(D_i\phi)(D_j\phi)^* \right]
\label{18}
\end{equation}
Here $i,j=1,2$.

One can see that the central charge (\ref{18}) coincides with
the topological charge (the quantized magnetic flux) of the
vortex configuration. Indeed, $Z$ can be rewritten in the form
\begin{equation}
Z = \int \partial_i{\cal V}^i d^2x
\label{18bis}
\end{equation}
where ${\cal V}^i$ is given by
\begin{equation}
{\cal V}^i =  \epsilon^{ij}\left(\frac{1}{2\,e}A_j +
\frac{i}{2\,e^2{\phi_0}^2}
{\phi}^*D_j\phi \right)
\label{18bisbis}
\end{equation}
so that, after using Stokes' theorem (and taking into account
that $D_i\phi \to 0$ at infinity)
\begin{equation}
Z = \frac{1}{e}  \oint A_idx^i = \frac{\pi n}{e}
\label{18bisbisbis}
\end{equation}
with $  n \in Z$ an integer characterizing the homotopy class
to which $A_{i}$ belongs.

Let us now introduce the  projector
\be {\cal P}_\pm = \frac12 (1 \mp \gamma_0) \label{24} \ee
and define
\be
{\cal Q}_\pm = {\cal P}_\pm  Q
\label{25}
\ee
Then, we project eq.(\ref{16}) with ${\cal P}$ and take the
trace getting
\be \{{\cal Q}_{\pm\,\alpha}, {\cal Q}^\dagger_{\pm\,\alpha} \}
= M \pm Z \label{esta} \ee
Taking the expectation value of (\ref{esta}) in an arbitrary
state and since the anticommutator of an operator with its
adjoint is a positive definite operator we conclude that
\begin{equation}
M \geq \vert  Z \vert
\label{20}
\end{equation}
or
\begin{equation}
M \geq \frac{\pi |n|}{e}
\label{21}
\end{equation}
which is the Bogomol'nyi bound for the vortex mass. For
positive (negative) values of $n$ the bound is attained only if
the state is annihilated by ${\cal Q}_+$ (${\cal Q}_-$),
\be {\cal Q}_\pm |BPS \rangle_\pm = 0 \label{carga-bps}
\ee
In terms of components this is equivalent to the condition
\be \left(Q_+ \pm i Q_-\right)|BPS \rangle_\pm = 0 \ee
In view of eqs. (\ref{14})-(\ref{15}), (\ref{24})-(\ref{25}),
equation \eqref{carga-bps} imply
\begin{align}
&B =  \mp \frac{e}{2} \,({\phi_0}^2 - |\phi|^2) \nonumber \\
& D_1\phi  =  \mp i \,D_2\phi
\label{27}
\end{align}
which are the BPS equations for the Abelian Higgs model. Due to
(\ref{21}), their solution also solves the static
Euler-Lagrange equations of motion. As in the kink case,
according to the choice of sign in the BPS equations, the
corresponding solution will break half of the supersymmetries.
Let us finally insist that the condition $\lambda e^2 =
\lambda/8$ necessary for this last fact, arises in the present
approach from the requirement of $N=2$ supersymmetry.

In order to connect supersymmetry with conditions on the stress
tensor, we will analyze the supercurrent-supercharge algebra in
the bosonic sector of the model. The relevant terms in the
supercharge $\bar Q$ and the spatial components ${\cal J}_i$ of
the supercurrent leading to (static) bosonic contributions are
\begin{align}
{\bar Q} &= \frac{1}{\sqrt{2}\,e\phi_0}\int d^2x \left\{
\overline{\Sigma}\left(-B - \frac{e}{2} \gamma^0 \left(|\phi|^2 -
\phi_0^2\right) \right) - {\bar \psi}\, \epsilon^{ij} \gamma^i D_j
\phi \right\} + \ldots \label{qui}\\
\nonumber\\
J^i &= \left[i B \, \epsilon^{ij}\, \gamma^j - \frac{e}{2}
\left(|\phi|^2 - \phi_0^2\right)\, \gamma^i\right]\, \Sigma - \left[
i\left(D_i\phi\right)^* - \epsilon^{ij}\, \left(D_j\phi\right)^*\,
\gamma^0\right] \psi + \ldots \nonumber\\
{\bar J}^i &= \overline{\Sigma}\, \left[- i B \, \epsilon^{ij}\,
\gamma^j - \frac{e}{2} \left(|\phi|^2 - \phi_0^2\right)\,
\gamma^i\right] + \bar \psi\, \left[ i\left(D_i\phi\right) +
\gamma^0\,\epsilon^{ij}\, \left(D_j\phi\right)\right] + \ldots
\nonumber\\
\label{jua}
\end{align}
where we have written the supercurrent $\mathcal{J}^i$ in the form
\be \mathcal{J}^i = \bar\eta_c\, J^i + {\bar J}^i \, \eta_c \ee
and ellipsis $\ldots$ indicate irrelevant terms which will be
ignored from here on.

From eqs.(\ref{qui})and (\ref{jua}) we find that
\begin{align}
\left\{ J^i_\alpha, \bar Q_\beta\right\} &= \frac{\sqrt{2}}{e\phi_0}
\left\{ \left( - \frac{1}{2}\, B^2 + \frac{e^2}{8} \left(|\phi|^2 -
\phi_0^2\right)^2 + \frac12\, |D_i\phi|^2 \right) \gamma^i_{\alpha
\beta} + \right.\\
&\;\;\; + \left. \frac12\,\left( \left( /\!\!\!\! D \phi\right)^*
D_i\phi -
\left(D_i\phi\right)^* /\!\!\!\!D\phi \right)_{\alpha \beta}
\right\}
\end{align}
and hence
\begin{align}
\text{Tr} \left( \gamma^i\, \left\{J^j, \bar Q\right\}\right) &=
\frac{2\, \sqrt{2} }{e\phi_0} \left\{ \left(B^2 -\frac{e}{2}^2
\left(|\phi|^2
- \phi_0^2\right)^2 - |D_k\phi|^2 \right) \delta^{ij} + \right.\\
&\left. + \left(D^i\phi\right)^*\, D^j \phi +
\left(D^j\phi\right)^*\, D^i \phi \right\}
\end{align}
Now, the r.h.s. is nothing but the symmetric stress tensor as
defined in (\ref{cuatres}), so that
\be
 T_{ij} = \frac{e\phi_0}{2\, \sqrt{2}}\,
\text{Tr} \left( \gamma_i\, \left\{J_j, \bar Q\right\}  \right)
\label{relacion}
 \ee
In particular we have
\begin{align}
%
&\left\{j^1_+ + i j^1_- , {\bar Q}_+ + i {\bar Q}_- \right\}  =
\frac{2 \sqrt{2}}{e\phi_0}
\left( T_{11} + i\, T_{21}\right)\\
&\left\{j^1_+ - i j^1_- , {\bar Q}_+ - i {\bar Q}_- \right\}  =
-\frac{2 \sqrt{2}}{e\phi_0}
\left( T_{11} - i\, T_{21}\right)\\
&\left\{j^2_+ + i j^2_- , {\bar Q}_+ + i {\bar Q}_- \right\}  =
\,\frac{2 \sqrt{2}}{e\phi_0}
\left( T_{12} - i\, T_{22}\right)\\
&\left\{j^2_+ - i j^2_- , {\bar Q}_+ - i {\bar Q}_- \right\}  =
-\,\frac{2 \sqrt{2}}{e\phi_0}
\left( T_{12} + i\, T_{22}\right)
\end{align}
But
\begin{align}
j^1_+\pm i j^1_- =& -  \left(B \mp \frac{e}{2}\,
(|\phi|^2 - \phi_0^2)\right)
\left(\Sigma_+\mp i\, \Sigma_-\right) +\\
&\;\;\;\;\;\; +\left(i\, (D_1\phi)^* \pm (D_2\phi)^*\right)
\left(\psi_+\pm i\,\psi_-\right)\\
j^2_+\pm i j^2_- =& \pm\, i\, \left(B \mp \frac{e}{2}\,
(|\phi|^2 - \phi_0^2)\right)
\left(\Sigma_+\mp i\, \Sigma_-\right) \pm\\
&\;\;\;\;\;\; +i\,\left(i\, (D_1\phi)^* \pm (D_2\phi)^*\right)
\left(\psi_+\pm i\,\psi_-\right)
\end{align}
Then, analogously to the kink case, either one has $\left({\bar
Q}_+ + i  {\bar Q}_-\right)|BPS\rangle_\pm = 0 $, or
$\left(j^1_++i  j^1_-\right)|BPS\rangle_\pm = 0$ and
$\left(j^2_++i  j^2_-\right)|BPS\rangle_\pm = 0$ (a similar
statement is valid for $\left({\bar Q}_+ - i  {\bar
Q}_-\right)$, $\left(j^1_+ - i  j^1_-\right)$, and $\left(j^2_+
- i\, j^2_-\right)$).

So we can write for BPS vortex states
\be
_\pm\langle BPS |T_{ij}|BPS\rangle_\pm = 0
\label{formuli}
\ee

At this point, it should be stressed that eq.(\ref{relacion})
from which the vanishing of the stress tensor components was
inferred is in general valid for other supersymmetric models in
which one can write
\be T_{ij} = {\cal N}_d \text{Tr} \left( \gamma_i\, \left\{J_j, \bar
Q\right\} + \gamma_j\, \left\{J_i, \bar Q\right\}\right)
\label{relacion2}
 \ee
where $T_{ij}$ is the symmetric stress-tensor and   ${\cal N}_d
$ a constant depending on the parameters of the specific model.
This is valid for the kink (eq. \eqref{shi}), for the vortex
(eq. \eqref{relacion}) but also for the  dyon, the instanton
taken as a soliton in $4+1$ dimensions, etc (see also
\cite{RvNW}-\cite{ShiV}). In particular, consider   the $3+1$
case, where the supercharge algebra for the $N=2$ Yang-Mills
theory takes the form
\be
\{Q^\alpha,\bar Q_{\,\beta}\} = -
\left(\gamma_\mu\right)^\alpha_{\,\beta} P_\mu
+ \left(\gamma_5\right)^\alpha_{\,\beta} U  +
i \delta^\alpha_{\,\beta} V
\label{mopn}
\ee
where $\alpha,\beta = 1,\ldots,4$ and the central charges $U$
and $V$ are surface integrals. If one takes as gauge group
$O(3)$ and breaks this symmetry to $U(1)$ by giving a non-zero
vacuum expectation value to the scalar field taken in the
adjoint, $U$ corresponds to the $U(1)$ magnetic charge and $V$
to the electric charge. A Bogomol'nyi bound can be then derived
from (\ref{mopn}),
\be M^2 \geq U^2 + V^2 \ee
and is saturated when the Bogomol'nyi-Prassad-Sommerfield
equations are satisfied. Now, one can see that
eq.(\ref{relacion2}) holds in this case   with the spatial
components of the supercurrent taking the form
\bea J_{i a} = {\rm Tr} \left( \sigma^{\mu\nu} F_{\mu\nu}
\gamma_i \Psi_a + \varepsilon_{ab} \not \!\!D \phi \gamma_i
\Psi_b \right)
\eea
This formula corresponds to a bosonic sector containing a gauge
field $A_\mu$  in the Lie algebra of $O(3)$ coupled to a Higgs
scalar $\phi$ in the adjoint (There is an additional
pseudoscalar field that should be put to zero to make contact
with the Georgi-Glashow model). Concerning the fermion sector,
$\Psi_a$ ($a=1,2$) are two Majorana fermions. Then, using
eq.(\ref{relacion2}) and proceeding as for the kink and the
vortex, one can see that eq.(\ref{formuli}) also holds  for the
Prasad-Sommerfield dyon. That is, the stress-tensor vanishes
for BPS dyons, a fact that can be trivially confirmed by
explicit computation of $T_{ij}$.
\\%

We have discussed in this note the relation between
supersymmetry and the vanishing of the stress tensor for
topological solitons in a variety of field theories in
different space-time dimensions. Each one of the elements in
this relation was already understood but our point was to show
how they could  be put together, by exploiting the relation
that exists in supersymmetric theories between the supercurrent
and the energy-momentum tensor.  In fact, this relation was
already underlying the analysis in ref.\cite{WO} where BPS
equations were derived from the relation between the
supercharge algebra and the energy momentum vector   $P_\mu =
\int d^3x T_{0\mu}$ which in the rest frame reduces to $P_0 =
M$.
\\%

Here, we have instead used the fact that, since the
supercurrent and the energy-momentum tensor belong to the same
multiplet, we can extend the analysis of the relation between
BPS states and supersymmetry to the spatial components of
$J_\mu$ and  $T_{\mu\nu}$. If we consider for example the
$d=3+1$ case  in the superfield framework, the linear $\theta$
component of the multiplet is the supercurrent and the $\theta
\bar \theta$ component corresponds the energy-momentum tensor
and they should then necessarily transform under supersymmetry
one into the other, \be \{J_\mu, \bar Q\} \propto \gamma^\nu
T_{\mu\nu} + \ldots \ee
Similar identities hold in other $d+1$ dimensional models. As
signaled above, eq.(\ref{relacion2}) leading to the connection
between supersymmetric BPS states with the condition $T_{ij}=0$
can be inferred from this formula. Now, as it is well known,
$T_{ij}$ gives the force $f_i$ acting in a unit volume of the
system. This, together with our result  means that, in general,
supersymmetry can guide the construction of non-interacting
solitons bosonic models of interest just by considering the
supersymmetric extension as a tool for identifying BPS states.

There are also possible applications of our observation in
supergravity models, in connection with stability of cosmic
strings \cite{ACEvBvP}-\cite{CSvP} and with the cosmological
constant problem \cite{BBS}-\cite{SN}. In particular, the
so-called dominant energy condition, $T_{00} \geq |T_{ij}|$,
valid for static spacetime, plays a centra role to establish a
connection between stability and the sign of the deficit angle
\cite{CSvP}. In this context it is natural to study
supergravity models with string-like BPS solutions in their
bosonic sector. An analysis based on the supercharge algebra
has been already presented \cite{ENS2} and it should be
worthwhile to study the problem from the point of view of
supercurrents presented here. We hope to report on these issues
in a future work.

\vspace{1 cm}

\noindent\underline{Acknowledgments}  We would like to thank
Carlos N\'u\~nez for helpful suggestions at the origin of this work.
This work was partially supported by   PIP6160-CONICET,  BID
1728OC/AR PICT20204-ANPCYT grants and by CIC and UNLP, Argentina.



\begin{thebibliography}{99}
\bibitem{Bogo} E.~B.~Bogomol'nyi, Sov.\ Jour.\ Nucl.\ Phys.\ B {\bf 24} (1976) 449.
%
\bibitem{dVS} H.~de Vega and F.~A.~Schaposnik, Phys.\ Rev.\ D {\bf 14} (1976) 1100.
%
\bibitem{DiVF} A.~Di Vecchia and S.~Ferrara, Nucl.\ Phys.\ B {\bf 130} (1977) 93.
%
\bibitem{WO} E.~Witten and D.~I.~Olive,
  Phys.\ Lett.\  B {\bf 78} (1978) 97.
  %
\bibitem{LLW}C.~k.~Lee, K.~M.~Lee and E.~J.~Weinberg,
  Phys.\ Lett.\  B {\bf 243} (1990) 105.
\bibitem{HS}
  Z.~Hlousek and D.~Spector,
  Nucl.\ Phys.\  B {\bf 370} (1992) 143; 
  Nucl.\ Phys.\  B {\bf 397} (1993) 173.
  \bibitem{ENS1} J.~D.~Edelstein, C.~N\'u\~nez and F.~Schaposnik,
  Phys.\ Lett.\  B {\bf 329}, 39 (1994).
\bibitem{ENS2}J.~D.~Edelstein, C.~N\'u\~nez and F.~A.~Schaposnik,
  Nucl.\ Phys.\  B {\bf 458}, 165 (1996); Phys.\ Lett.\
   B {\bf 375}, 163 (1996).
\bibitem{RvNW} A.~Rebhan, P.~van Nieuwenhuizen and R.~Wimmer,
Braz.\ J.\ Phys.\  {\bf 34} (2004) 1273.
%
\bibitem{SY} M.~Shifman and A.~Yung,
  Rev.\ Mod.\ Phys.\  {\bf 79} (2007) 1139.
%
\bibitem{Sohnius}  M.~F.~Sohnius,
Phys.\ Rept.\  {\bf 128} (1985) 39.
%
\bibitem{Manton} N.~S.~Manton,
arXiv:0809.2891 [hep-th].
%
\bibitem{Abri} A.~Abrikosov, Sov.\ Phys.\ JETP {\bf 32} (1957) 1442.
[Reprinted in {\it Solitons and Particles}, Eds. C.~Rebbi and
G.~Soliani, World Scientific, Singapore, 1984.
%
\bibitem{dAdV}  A.~D'Adda and P.~Di Vecchia,
Phys.\ Lett.\  B {\bf 73} (1978) 162.
%
\bibitem{ShiV} M.~A.~Shifman and A.~I.~Vainshtein, {\it ITEP
    Lectures in Particle Physics and Field Theory. Ed. by M.~A.~Shifman.
    World Scientific,  Vol. 2, Singapore,
1999}.
  \bibitem{ACEvBvP}
  A.~Achucarro, A.~Celi, M.~Esole, J.~Van den Bergh and A.~Van Proeyen,
  JHEP {\bf 0601} (2006) 102
  [arXiv:hep-th/0511001].

 \bibitem{CSvP} A.~Collinucci, P.~Smyth and A.~Van Proeyen,
  JHEP {\bf 0702} (2007) 060.
  \bibitem{BBS}  K.~Becker, M.~Becker and A.~Strominger,
  Phys.\ Rev.\  D {\bf 51}, 6603 (1995).
  \bibitem{SN}
  S.~Nobbenhuis,, Found.\ Phys.\  {\bf 36} (2006) 613;
  arXiv:gr-qc/0609011.




\end{thebibliography}
\end{document}